\documentclass[a4paper]{jpconf}
\usepackage{graphicx}
\newcommand{\bea}{\begin{eqnarray}}
\newcommand{\eea}{\end{eqnarray}}

\newcommand{\deltaunodue}{\mbox{$\Delta m_{21}^2 $}}
\newcommand{\deltaunotre}{\mbox{$\Delta m_{31}^2 $}}

\newcommand{\tetaot}{\mbox{$\theta_{13}$}}

\newcommand{\nova}{\mbox{NO$\nu$A} }

\newcommand{\beq}{\begin{eqnarray}}
\newcommand{\eeq}{\end{eqnarray}}
\begin{document}
\begin{flushright}
{\tt hep-ph/0609031}\\
{FERMILAB-CONF-06-282-T}
\end{flushright}
\title{Getting the most from NO$\nu$A and T2K}

\author{Olga Mena }

\address{Theoretical Physics Department,
Fermi National Accelerator Laboratory, 
P.O.Box 500, Batavia, IL 60510, USA}

\ead{omena@fnal.gov}

\begin{abstract}
The determination of the ordering of the neutrino masses (the hierarchy) is probably a crucial prerequisite to understand the origin of lepton masses and mixings and to establish their relationship to the analogous properties in the quark sector. In this talk, we follow an alternative strategy to the usual neutrino--antineutrino comparison: we exploit the combination of the neutrino-only data from the \nova and  the T2K experiments by performing these two off-axis experiments at different distances but at the same $\langle E \rangle/L$, $\langle E\rangle$ being the mean neutrino energy and $L$ the baseline. This would require a minor adjustment to the proposed off-axis angle  for one or both of the proposed
experiments.   
\end{abstract}

\section{Introduction}
During the last several years the physics of neutrinos has achieved a
remarkable progress. The experiments with solar~\cite{sol,SKsolar,SNO1,SNO2,SNO3,SNOsalt},  atmospheric~\cite{SKatm}, reactor~\cite{KamLAND}, and also
long-baseline accelerator~\cite{K2K,MINOSRECENT} neutrinos, have provided compelling evidence for the existence of neutrino oscillations, implying non zero neutrino
masses. The data quoted above require two large mixing angles
($\theta_{12}$ and $\theta_{23}$) and may involve a small third one ($\theta_{13}$) 
in the neutrino mixing matrix and two mass squared differences,
$\Delta m_{ji}^{2} \equiv m_j^2 -m_i^2$, with $m_{j,i}$ the neutrino
masses, one driving the atmospheric ($\deltaunotre$) and the other one the solar ($\deltaunodue$) neutrino oscillations. The mixing
angles $\theta_{12}$ and $\theta_{23}$ control the solar and the atmospheric neutrino oscillations, while $\theta_{13}$ is the
angle limited by the data from the CHOOZ and Palo Verde reactor
experiments~\cite{CHOOZ,PaloV}.

The Super-Kamiokande~\cite{SKatm} and K2K~\cite{K2K} data are well
described in terms of dominant $\nu_{\mu} \rightarrow \nu_{\tau}$
($\bar{\nu}_{\mu} \rightarrow \bar{\nu}_{\tau}$) vacuum
oscillations. The MINOS Collaboration has recently reported their first neutrino oscillation results from $1.27 \times 10^{20}$ protons on target exposure of the MINOS far detector~\cite{MINOSRECENT}. A recent global fit~\cite{thomas} (see also Ref.~\cite{newfit}) provides the following $3 \sigma$ allowed ranges for the atmospheric mixing parameters:
\beq 
\label{eq:range}|\deltaunotre| =(1.9 - 3.2)\times10^{-3}{\rm eV^2},~~~~
0.34<\sin^2\theta_{23}<0.68~.
\eeq
The sign of $\deltaunotre$, sign$(\deltaunotre)$, 
cannot be determined with the existing data. The two possibilities,
$\deltaunotre > 0$ or $\deltaunotre < 0$, correspond to two different
types of neutrino mass ordering: normal hierarchy and inverted hierarchy. 
In addition, information on the octant in which $\theta_{23}$ lies, if $\sin^22\theta_{23} \neq 1$, is beyond the reach of present experiments. 

The 2-neutrino oscillation analysis of the solar neutrino data, in combination
with the KamLAND spectrum data~\cite{KL766}, shows that the solar neutrino oscillation parameters lie in the low-LMA (Large Mixing Angle) region, with best fit values~\cite{thomas} $\deltaunodue =7.9 \times 10^{-5}~{\rm eV^2}$ and $\sin^2 \theta_{12} =0.30$.


A combined 3-neutrino oscillation analysis of the solar, atmospheric,
reactor and long-baseline neutrino data~\cite{thomas} constrains the third mixing angle to be $\sin^2\theta_{13} < 0.041$ at the $3\sigma$ C.L.

The future goals in the study of neutrino properties will be to
measure precisely the already known oscillation parameters 
and to obtain information on the unknown ones, namely $\theta_{13}$,
the CP--violating phase $\delta$ and the neutrino mass
hierarchy (or equivalently sign$(\deltaunotre)$). In this talk~\cite{prep}, we concentrate on the extraction of the neutrino mass hierarchy 
by combining the Phase I (neutrino-data only) of the long-baseline $\nu_e$ appearance experiments  T2K~\cite{T2K} and NO$\nu$A~\cite{newNOvA}, both exploiting the off-axis technique~\cite{adamoff}. For our analysis, unless otherwise stated, we will use a representative value of $|\deltaunotre| = 2.4 \times 10^{-3} \
\rm{eV}^2$ and $\sin^2 2 \theta_{23}=1$. 
For the solar oscillation parameters $\deltaunodue$ and $\theta_{12}$, we will use the best fit values
quoted in this introductory section. 

\section{Formalism}
\label{formalism}
The mixing angle $\theta_{13}$ controls $\nu_\mu \rightarrow
\nu_e$ and $\bar{\nu}_\mu \rightarrow \bar{\nu}_e$ conversions in
long-baseline $\nu_e$ appearance experiments and the $\bar{\nu}_e$ disappearance in short-baseline reactor experiments.  Present and future reactor neutrino experiments~\cite{futurereactors}, conventional neutrino beams and future long baseline neutrino experiments could measure, or set a stronger limit on, $\theta_{13}$. Therefore, with the possibility of the first measurement of $\theta_{13}$ being made by a 1-to 2-km baseline reactor experiment, the long-baseline off-axis $\nu_e$ appearance experiments, T2K~\cite{T2K} and NO$\nu$A~\cite{newNOvA},  need to adjust their focus to emphasize other physics topics.  
The most important of these questions is the form of the mass hierarchy, normal
versus inverted and the measurement of leptonic CP violation, which in a three neutrino oscillation framework is directly related to the existence of a CKM-like CP-phase, $\delta$.
 Consider the probability $P (\nu_\mu \to \nu_e) $ in the context of three-neutrino mixing in the presence of matter~\cite{matter}, represented by the matter parameter $a$, defined as $a \equiv G_{F} n_e/\sqrt{2}$, where $n_e$ is the average electron number density over the baseline, taken to be constant throughout the present study. Defining $\Delta_{ij} \equiv \frac{\Delta m^2_{ij} L}{4 E}$, a convenient and
precise approximation is obtained by expanding to second order 
in the following small parameters: 
$\tetaot$, $\Delta_{21}/\Delta_{32}$, $\Delta_{21}/aL$ and $\Delta_{21}$. 
The result is (details of the calculation can be found in Ref.~\cite{CDGGCHMR00}, see also Ref.~\cite{3probnew})~\footnote{The author would like to thank S.~Parke for the shorter version of the oscillation probability below.}: 
\bea
P_{\nu_ \mu \nu_e}\simeq \left|\sin \theta_{23}\sin 2 \theta_{13}
\left(\frac{\Delta_{31}}{\Delta_{31}-aL}\right)\sin(\Delta_{31}-aL)e^{-i(\Delta_{32}+\delta)}+\cos\theta_{23}\sin 2 \theta_{12}\left(\frac{\Delta_{21}}{aL}\right) \sin \left( aL \right) \right|^2 
\label{eq:probappr}
\eea
where $L$ is the baseline and $a\to -a$, $\delta\to -\delta$ for $P_{\bar \nu_\mu \bar \nu_e}$.  
Suppose $P_{\nu_ \mu \nu_e} < P_{\bar \nu_\mu \bar \nu_e}$: in vacuum, this implies CP violation. On the other hand, in matter, this implies CP violation only for the normal hierarchy but not necessarily for the inverted hierarchy around the first oscillation maximum. The different index of refraction for neutrinos and antineutrinos induces differences in the $\nu$, $\bar{\nu}$ propagation that could be misinterpreted as CP violation~\cite{matterosc}. 
Typically, the proposed long baseline neutrino oscillation experiments have a single detector and plan to run with the beam in two different modes, neutrinos and antineutrinos. In principle, by comparing the probability of neutrino and antineutrino
flavor conversion, the values of the CP--violating phase $\delta$ and
of sign$(\deltaunotre)$ could be extracted. However, different sets of values
of CP--conserving and violating parameters, ($\theta_{13}$, $\theta_{23}$,
$\delta$, sign$(\deltaunotre)$), lead to the same probabilities of
neutrino and antineutrino conversion and provide a good description of
the data at the same confidence level. This problem is known as the
problem of degeneracies in the neutrino parameter
space~\cite{FL96,BCGGCHM01,MN01,BMWdeg,deg} and severely affects the
sensitivities to these parameters in future long-baseline experiments.
Many strategies have been advocated to resolve this issue. Some of the
degeneracies might be eliminated with sufficient energy or baseline
spectral information. In practice, statistical errors
and realistic efficiencies and backgrounds limit considerably the
capabilities of this method. Another
detector~\cite{BCGGCHM01,MN97,silver,BMW02off,twodetect} or
the combination with another
experiment~\cite{noi,HLW02,MNP03,BMW02,otherexp,mp2,HMS05,M05,huber1,huber2} would,
thus, be necessary.

The use of only a neutrino beam could help in resolving the type of
hierarchy when two different long-baselines are considered~\cite{HLW02,MNP03,SN1,SN2}. It was shown in ref.~\cite{MNP03} that if the $\langle E\rangle/L$ for the two
different experiments is approximately the same then the allowed regions
for the two hierarchies are disconnected and thus this method for determining
the hierarchy is free of degeneracies.
Naively, we can understand this 
method in the following way for $\sin^2 2 \theta_{13}>0.01$: 
assume that matter effects are negligible for the short baseline, 
then at the same $\langle E\rangle/L$, if the oscillation probability 
at the long baseline is larger than the oscillation probability 
at the short baseline, one can conclude that the hierarchy is normal, since 
matter effects enhance the neutrino oscillation probabilities for 
the normal hierarchy.
For the inverted hierarchy the oscillation probability 
for the long baseline
is suppressed relative to the short baseline

\section{ Our strategy: only neutrino running and two detectors}
\label{strategy}

Following the line of thought developed by Minakata, Nunokawa and Parke~\cite{MNP03}, we exploit the neutrino data from two experiments at different distances and at different off-axis locations~\cite{prep}. The off-axis location of the detectors and the baseline must be chosen such that the $\langle E \rangle/L$ is the same for the two experiments. Here we explain the advantages of such an strategy versus the commonly exploited neutrino-antineutrino comparison.

 Suppose we compute the oscillation probabilities $P_{\nu_ \mu \nu_e}$ and $P_{\bar \nu_\mu \bar \nu_e}$ for a given set of oscillation parameters and the CP-phase $\delta$ is varied between $0$ and $2 \pi$: we obtain a closed CP trajectory (an ellipse) in the bi--probability space of neutrino and antineutrino conversion~\cite{MN01}. In general, the ellipses overlap for a large fraction of values of the CP--phase $\delta$ for every allowed value of $\sin^2 2
\theta_{13}$.  This indicates that, generically, a measurement of the probability of conversion for neutrinos and antineutrinos cannot uniquely determine
the type of hierarchy in a single experiment. This makes the determination of sign$(\deltaunotre)$ extremely difficult, i.~e., the sign$(\deltaunotre)$-extraction is not free of degeneracies.

In the case of bi--probability plots of neutrino--neutrino conversions at different distances (which will be referred as near (N) and far (F)), the overlap of the two bands, which implies the presence of a degeneracy of the type of hierarchy with other parameters, is controlled by the slope and the width of the
bands. Using the fact that matter effects are small ($aL\ll\Delta_{13}$), we can perform a perturbative expansion and assuming that the $\langle E \rangle/L$ of the near and far experiments is the same, at first order, the ratio of the slopes reads~\cite{MNP03}
\beq
\frac{\alpha_+}{\alpha_-} \simeq
1 +  4 \left( a_{\rm N} L_{\rm N} - a_{\rm F} L_{\rm F} \right)\left( \frac{1}{\Delta_{31}} - \frac{1}{\tan(\Delta_{31})} \right)~,
\label{eq:ratioapp}
\eeq
where $\alpha_+$ and $\alpha_-$ are the slopes for normal and inverted
hierarchies, and $a_{\rm F}$ and $a_{\rm N}$ are the matter parameters for the two experiments. The separation among the ellipses for the two hierarchies increases as the matter parameter times the path length for the two experiments does. The width of the ellipses is crucial: even when the separation between the central axes of the two regions is substantial, unless the ratio $\langle E \rangle/L$ is kept close to constant, the width of the ellipses will grow rapidly and the ellipses will overlap. 
Consequently, we have to satisfy two conditions in order to optimize the determination of the neutrino mass hierarchy: (a) maximize the difference in the factor $a L$ for both experiments and  (b) minimize the ellipses width by performing the two experiments at the same $\langle E \rangle/L$. 

The most promising way to optimize the sensitivity to the hierarchy with relatively near term data is therefore to focus on the neutrino running mode and to exploit the Phase I data of the long-baseline off-axis $\nu_e$ appearance experiments, T2K  and NO$\nu$A. T2K utilizes a steerable neutrino beam from JHF and Super-Kamiokande and/or Hyper-Kamiokande as the far detector. The beam will peak at $0.65$ GeV by placing the detector off-axis by an angle of $2.5^\circ$ at $295$ km. 
NO$\nu$A proposes to use the Fermilab NuMI beam with a baseline of 
$810$~km  with a $30$ kton low density tracking calorimeter with an efficiency of $24\%$. Such a detector would be located $12$~km off-axis, resulting in a mean neutrino energy of $2$ GeV. 
While for the T2K experiment matter effects are non negligible, albeit small~\cite{matter_t2k}, matter effects are quite significant for NO$\nu$A. Therefore, the condition (a) is satisfied, since $(aL)_{\textrm{\nova}} \simeq 3 (aL)_{\textrm{T2K}}$ . What about the condition (b)? A back-of-the-envelope calculation indicates that the current off-axis detector locations are not such that $\langle E \rangle/L$ of the two experiments is the same. However, by placing the detector(s) in slightly different off-axis location(s),
 one can manage the $\langle E \rangle/L$ of the two experiments to be exactly the same. This neutrino-data strategy  would only need half of the time of data taking (because we avoid the antineutrino running), when compared to the standard one (i.e. running in neutrinos and antineutrinos at a fixed energy, $E$, and baseline, $L$).

\section{Optimizing the \nova and T2K detector locations}
\label{optimal}
In this section we present what could be achieved if \nova and T2K setups are carefully chosen, focusing on the physics potential of the combination of their future data. We define the Phase I of the experiments as follows. For the T2K experiment, we consider 5 years of neutrino running and SK as the far detector with a fiducial mass of $22.5$ kton and $70\%$ detection efficiencies. For the \nova experiment, we assume $6.5\times 10^{20}$ protons on target per year, 5 years of neutrino running and the detector described in the previous section.
\begin{figure}[t]
\begin{center}
\begin{tabular}{ll}
\includegraphics[width=3in]{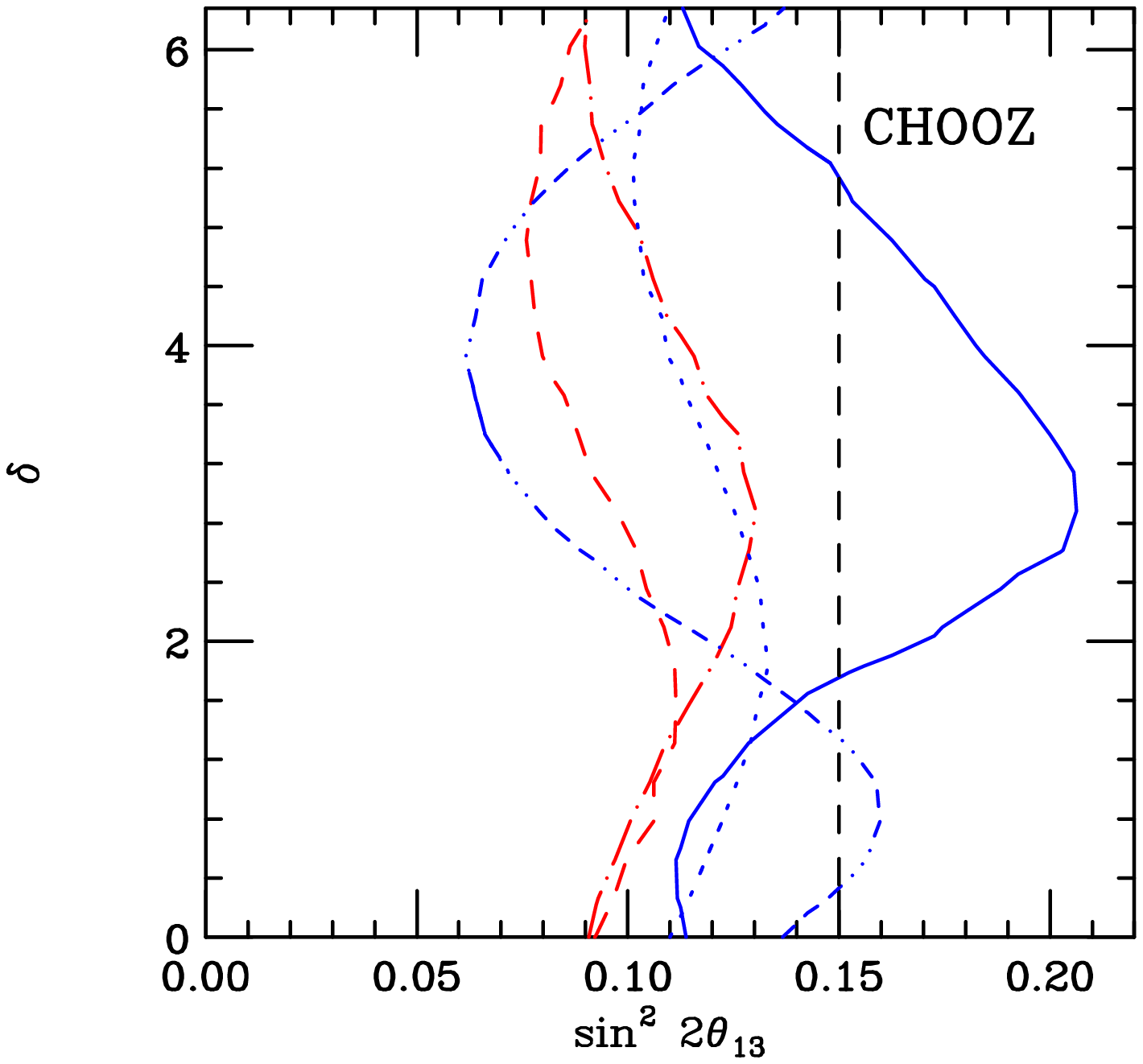}&\hskip 0.cm
\includegraphics[width=3in]{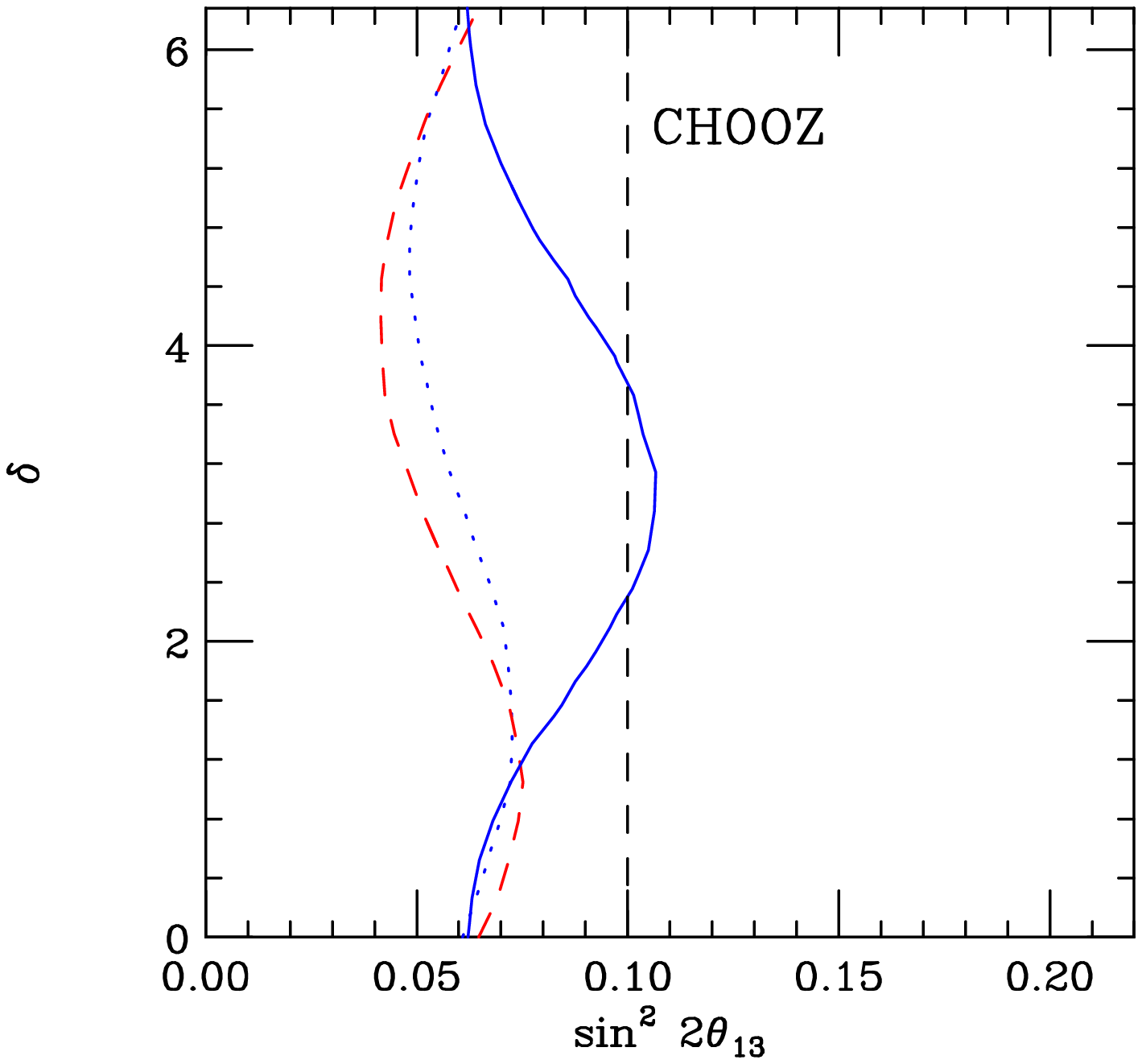}\\
\hskip 2.truecm
{\small (a) $\Delta m^2_{31} = 2.4 \times 10^{-3}$~eV$^2$}            &
\hskip 2.truecm
{\small (b) $\Delta m^2_{31} = 3 \times 10^{-3}$~eV$^2$}       \\  
\end{tabular}
\end{center}
\caption[]{\textit{(a) $90\%$ CL hierarchy resolution (2 d.o.f) for different possible combinations: the \textit{default} one (T2K at an off-axis angle of $2.5^\circ$ and \nova far detector at $12$~km off-axis, in solid blue),  T2K at an off-axis angle of $2.5^\circ$ and \nova far detector at $13$~km off-axis (long dash-dot red curve),  at $14$~km off-axis (short dashed red curve), at $16$~km off-axis (three dots-three dashes blue curve) and  T2K at an off-axis angle of $2^\circ$ and \nova far detector at $12$~km off-axis (dotted blue curve). (b) The same as (a) but assuming that $\Delta m^2_{31}= 3.0 \times 10^{-3} \ eV^2$ and only for the three most representative combinations.}}\label{fig:exclusion}
\end{figure}

We summarize the results in Figs.~(\ref{fig:exclusion}), where we present the exclusion plots in the ($\sin^2 2\theta_{13}, \delta$) plane for a measurement of the hierarchy at the $90\%$ CL for the several possible combinations, assuming that nature's solution is the normal hierarchy and $\Delta m^2_{31}=2.4\times 10^{-3}$~eV$^2$ (left panel) and $\Delta m^2_{31}=3 \times 10^{-3}$~eV$^2$ (right panel) (in light of the recent MINOS results, we explore here the impact of a larger  $\Delta m^2_{31}$). We show as well the corresponding CHOOZ bound for $\sin^2 2 \theta_{13}$. A larger value of $\Delta m^2_{31}$ implies more statistics and consequently a sensitivity improvement: see Fig.~(\ref{fig:exclusion}) (b), where for the sake of illustration only the three most representative configurations are shown.

If both T2K and \nova run in their \textit{default} configurations the combination of their future Phase I data (only neutrinos) will not contribute much  to our knowledge of the neutrino sector, see the solid blue line in Figs.~(\ref{fig:exclusion}). If we fix the T2K off-axis location to its \textit{default} value of $2.5^\circ$ but we change the location of the \nova detector to $14$~km the improvement is quite remarkable, see the short dashed red line in  Figs.~(\ref{fig:exclusion}): the sensitivity to the mass hierarchy has a milder dependence on the CP-phase $\delta$ once that the $\langle E \rangle/L$ of the two experiments is chosen to be the same. The best sensitivity to the hierarchy extraction is clearly achieved 
when the \nova experiment is at $14$~km off-axis and the T2K off-axis angle is 
the \textit{default} one. If the T2K off-axis angle is slightly modified to $2^\circ$, see the dotted lines in Figs.~(\ref{fig:exclusion}) it would be possible to reproduce the results from the combination of the data from T2K located at $2.5^\circ$ off-axis and the \nova detector placed at $13$~km off-axis.

The combination of data from an upgraded phase (Phase II) of the T2K and/or \nova experiments (by increasing the proton luminosities, the years of neutrino running and/or the mass of the far detectors) will obviously increase the statistics and will shift the sensitivity curves depicted in  Fig.~(\ref{fig:exclusion}) (a), similarly to the effect of increasing $\Delta m^2_{31}$.

If the nature's choice for the neutrino mass ordering is the inverted hierarchy, the  sensitivity curves depicted in  Fig.~(\ref{fig:exclusion}) (a) will be shifted but in the opposite direction, making the case for the Phase II of both experiments stronger, especially if $\Delta m^2_{31}=2.4 \times 10^{-3}$~eV$^2$.

\section{Conclusions}
\label{conclusions}

The most promising way to extract the neutrino mass hierarchy  is to make use of the matter effects and exploit the neutrino data from two near-term long baseline $\nu_e$ appearance experiments performed at the same $\langle E \rangle/L$, provided $\sin ^2 2 \theta_{13}$ is within their sensitivity range or within the sensitivity range of the next-generation $\bar{\nu}_e$ disappearance reactor neutrino experiments. Such a possibility could be provided by the combination of the data from the Phase I of the T2K and \nova experiments. We conclude that the optimal configuration for these experiments would be $14$~km off-axis for the \nova far detector and $2.5^\circ$ off-axis for the T2K experiment. The combination of their expected results could provide a $90\%$ confidence level resolution of the neutrino mass hierarchy if $\sin^2 2\theta_{13} > 0.11$ (for $\Delta m^2_{31}=2.4\times 10^{-3}$~eV$^2$) or if $\sin^2 2\theta_{13} > 0.07$ (for $\Delta m^2_{31}=3 \times 10^{-3}$~eV$^2$).
A modest upgraded next Phase of both \nova and T2K experiments (by increasing a factor of five their expected Phase I statistics) could shift the $90\%$ CL limits quoted above to $\sin^2 2\theta_{13} > 0.03$ (for $\Delta m^2_{31}=2.4\times 10^{-3}$~eV$^2$) and to $\sin^2 2\theta_{13} > 0.025$ (for $\Delta m^2_{31}=3 \times 10^{-3}$~eV$^2$).

\ack
The material presented here is based on work developed in collaboration with H.~Nunokawa, S.~Palomares-Ruiz, S.~Parke and S.~Pascoli. Fermilab is operated by URA under DOE contract DE-AC02-76CH03000.

\end{document}